\documentclass[twocolumn,showpacs,preprintnumbers,prd,superscriptaddress,
groupaddress,nofootinbib]{revtex4}

\usepackage{amsmath}
\usepackage{amsfonts}
\usepackage{graphicx}
\usepackage{natbib}

\def\sqr#1#2{{\vcenter{\hrule height.#2pt\hbox{\vrule width.#2pt height#1pt
\kern#1pt \vrule width.#2pt}\hrule height.#2pt}}}
\def\square{\mathchoice\sqr64\sqr64\sqr{4.2}3\sqr{3.0}3}

\begin{document}
\title{A spherical scalar--tensor galaxy model}
\author{Jorge L. Cervantes-Cota}
\author{Mario A. Rodr\'iguez-Meza}
\affiliation{Depto. de F\'{\i}sica, Instituto Nacional de
Investigaciones Nucleares, Apdo. Postal 18-1027, M\'{e}xico D.F. 11801,
M\'{e}xico.}
\author{Dar\'{\i}o N\'u\~nez}
\affiliation{Instituto de Ciencias Nucleares,  Universidad Nacional
Aut\'onoma de M\'exico, A.P. 70-543,  04510 M\'exico D.F.,
M\'exico}

\collaboration{Instituto Avanzado de Cosmolog\'{\i}a, IAC} \noaffiliation

\begin{abstract}
We build a spherical halo model for galaxies using a general
scalar--tensor theory of gravity in its Newtonian limit.   The
scalar field is described by a time-independent Klein--Gordon equation with a source 
that is coupled to the standard Poisson equation of Newtonian
gravity. Our model, by construction, fits both the observed rotation
velocities of stars in spirals and a typical luminosity profile. As
a result, the form of the new Newtonian potential, the scalar field, 
and dark matter distribution in a galaxy are determined. Taking into account the 
constraints for the fundamental parameters of the theory $(\lambda, \, \alpha)$,
we analyze the influence of the scalar field in the dark matter distribution, resulting 
in shallow density  profiles in galactic centers.
\end{abstract}

\date{\today}
\preprint{}
\maketitle

\section{Introduction}
The recent indirect observational evidence of dark matter (DM) and dark energy in the Universe
\cite{Ri98,Pe99,deB00,Ri01,Pe02,Ef02,Be03,Sp07,Hi08} has motivated the
study of new cosmological and astrophysical scenarios that can
encompass these observations.  At a cosmological level, the
quintessence scenario  \cite{CaDaSt98,Bo00,Am01} provides fittings
to the present accelerated expansion rate of the Universe and should be
consistent with the other above--mentioned observations. Typically,
quintessence introduces new fields that have their origin in theories that attempt to unify
all forces of nature (strings, braneworlds).  These unification
schemes result in extensions to General Relativity (GR) which
determine a new dynamics. Scalar--tensor theories (STT) of gravity
are examples of effective theories that stem from such unifying
schemes \cite{Gr88,GaVe03}.  As one may suspect, in addition to
cosmological consequences, such theoretical extensions also predict
local, astrophysical effects.  Traditionally, to understand the dynamics of a galaxy,  a DM profile  has 
been introduced \cite{OsPe73} or, alternatively, a modification to the Newtonian 
gravitational law, e.g. via Yukawa couplings \cite{Sa84},  or  even modifications to the Newtonian
motion law, such as MOND \cite{MOND1,MOND2}.  Within the first approach, several DM candidates  
have been proposed, including a scalar field (SF) as DM itself \cite{varios-gm04,MaGu01}.
The second approach has been used to obtain flat rotation curves  in spirals via STT,  without using DM halos.   
Adjusting rotation curve profiles using STT implies that the strength of the Yukawa coupling, $\alpha$, has 
to be negative, leading to a phantomlike nonminimally coupled field. Moreover,  the adjustment of 
different rotation curves of specific galaxies, points to different values for the range parameter of the 
theory, $\lambda$ \cite{Ag01}, implying a mass spectrum for the fundamental theory at hand.   The third 
approach solves flat rotation curves dynamics in spirals, but fails to fully understand cluster dynamics. This latter 
approach will not be considered here.  Our model is a combination of the first two approaches mentioned above: 
we use both a DM halo and a SF nonminimally coupled to GR, as long as these two elements could 
simultaneously play a role in the galactic dynamics, and mitigate the constraints imposed on STT 
parameters, when using that theory alone.   
 
Following this line of thought,  we have recently studied \cite{RoCe04,RoCePeTlCa05} the 
influence that STT have at galactic scales, see also \cite{BeDe07}. We considered the Newtonian 
limit of STT and computed potential--density pairs of a spherical galaxy in which 
NFW's \cite{Na96-97} and Dehnen's \cite{De93} density profiles
were used. We also computed some other relevant observational
quantities (rotation curves, dispersions), by which we accounted for
the influence of the STT scalar fields in the Galaxy. 
Such influence is characterized by the two parameters ($\lambda, \alpha$)
of STT, see below. In this work we use this formalism to 
build a galactic model that is, by construction, consistent with the
measured rotation curves of stars and with some luminosity profile.
As a result, the form of the Newtonian potential is exactly solved, and the SF  and DM
distribution in a galaxy are numerically computed,   
and their specific features depend on the fundamental parameters  $(\lambda, \,
\alpha)$ of the STT.  For the values of the parameter space analyzed, the resulting DM has  
shallow profiles near the center.

Some other models such as the gravitational suppression  hypothesis \cite{PiMa03} have been put forward, in which 
a Yukawa term is added in the Newtonian potential. Recently,   \cite{FrSa07} analyzed  the rotation curves best 
fit in this model, in which a NFW profile is used. They concluded that this hypothesis 
does not fit  several rotational curves of spirals and, hence, does not solve the core/cuspy problem of DM in the 
center of galaxies. Arguments in favor of a cusp like center can be found in Refs.  \cite{Po03, Na04, Ha04, MaGe04, Po04},  however  
recently more evidence has emerged  favoring  a corelike galactic center  
\cite{Sa01,VaSw01,We01,BiEv01,SwMaVaBa03,WedeWa03,Ge04,DoGeSa04,Sa04,Ge05, Si05, Ge07}. In \cite{SpGiHa05} 
 a way can be found to reconcile both approaches.

The present work is organized as follows: in the next section we give a brief description
of the STT Newtonian approximation, where the SF  background value is set
to have the usual gravitational constant value at small distances, $r\ll \lambda$. In Sec. \ref{galactic-model},
we build a galactic model by giving the rotation curves and baryon density profile, and in Sec. \ref{spherical-sol} we analyze the 
influence of the pair of parameters ($\lambda, \alpha$) on the SF and DM density distributions. Finally, in the last 
section we discuss the results and present our conclusions.


\section{The Newtonian Approximation of STT} \label{newapprox}
A typical STT is given by the following Lagrangian \cite{BrDi61,Wa70}:
\begin{equation}
\label{SFLagrangian}
    {\cal L} = \frac{\sqrt{-g}}{16\pi} \left[
    -\phi R + \frac{\omega(\phi)}{\phi} (\partial \phi)^2 - V(\phi)
    \right] + {\cal L}_M(g_{\mu\nu}) \; ,
\end{equation}
where $g_{\mu\nu}$ is the metric, $\phi$ is a SF, and $\omega(\phi)$ and $V(\phi)$ are arbitrary functions of it.   ${\cal L}_M(g_{\mu\nu})$ is the
 matter Lagrangian. From Eq. (\ref{SFLagrangian})  one obtains the gravity and SF equations. Thus, the
 gravitational equation is

\begin{eqnarray}
\label{Einsteineqn}
R_{\mu\nu} - \frac{1}{2}g_{\mu\nu}R  &=& \frac{1}{\phi}  \left[ 8\pi T_{\mu\nu} + \frac{1}{2} V(\phi) g_{\mu\nu} 
+ \frac{\omega(\phi)}{\phi}\partial_{\mu} \phi\ \partial_{\nu} \phi  
\right. \nonumber \\
&& \left.   -\frac{1}{2}\frac{\omega(\phi)}{\phi}(\partial_{\mu} \phi)^2
    g_{\mu\nu}+{\phi_{;\mu\nu}}-g_{\mu\nu}\square\,\phi
\right]. 
\end{eqnarray}
The SF Klein-Gordon equation is 
\begin{equation}
\label{SFparteqn}
    \square\,\phi + \frac{\phi V'-2V}{3+2\omega}=
    \frac{1}{3+2\omega}\left[8\pi T -{\omega}' (\partial \phi)^2\right]  \, ,  
\end{equation}
where a prime (') denotes the derivative with respect to SF.

 In accordance with the Newtonian approximation, gravity and SF are
 weak. Then, we expect to have small deviations of the SF 
 around the background field.  Assuming also that the velocities of stars and DM particles 
 are nonrelativistic, we perform the expansion of the field equations around the
background quantities $\langle\phi\rangle$ and $\eta_{\mu\nu}$, i.e., $g_{\mu\nu} = \eta_{\mu\nu} + h_{\mu\nu}$ and   
$\phi=\langle \phi \rangle + \delta \phi $.

The equations governing the weak energy (Newtonian) limit of STT are well known 
\cite{No70,He91,Wi93,RoCe04, CeRoGaKl07} and written here in physical units ($h\neq 1, c\neq1$)
\begin{eqnarray}
\frac{1}{2} \nabla^2 h_{00} &=& \frac{G{_N}}{(1+\alpha) c^{2}} \, \left[
4\pi \rho - \frac{1}{2} \nabla^2 \delta \phi \right] \; ,
\label{pares_eq_h00}\\
  \nabla^2 \delta \phi - \left(\frac{m c}{h}\right)^{2} \delta \phi &=&
-  8\pi \alpha\rho \; , \label{pares_eq_phibar}
\end{eqnarray}
where the background value is chosen such as $\langle \phi \rangle = (1+\alpha) c^2/G_{N}$, 
a choice that sets the effective Newtonian constant to the one locally 
observed, see \cite{CeRoGaKl07, CeRoNu07} for a detailed discussion. Eqs. (\ref{pares_eq_h00}) and 
(\ref{pares_eq_phibar}) re\-pre\-sent the Newtonian limit of a set 
of STT that are distinguished by the effective square mass 
$m^2 
\equiv 
\alpha (\langle \phi \rangle V''_{\phi = \langle \phi \rangle} - V'_{\phi = \langle \phi \rangle})  
- 2 \alpha^2 \omega'_{\phi = \langle \phi \rangle} (\langle \phi \rangle V'_{\phi = \langle \phi \rangle} 
- 2 V_{\phi = \langle \phi \rangle})$ 
and 
$\alpha \equiv 1 / (3 + 2\omega(\phi))\mid_{\phi= \langle \phi \rangle}$; $\omega(\phi)$ 
is a generalization of the Brans--Dicke parameter \cite{BrDi61}.  

In the above expansion we have set the cosmological constant equal
to zero since within galactic scales its influence is negligible.
This is because the average density in a galaxy is much larger than
a cosmological constant that is compatible with observations.  Thus,
we only consider the influence of luminous and dark matter. These
matter components gravitate in accordance  with the
modified--Newtonian theory determined by Eqs. (\ref{pares_eq_h00})
and \ (\ref{pares_eq_phibar}). The latter is a Klein--Gordon
equation which contains the boson field of mass $m$, whose Compton
wavelength ($\lambda = h/m c$) implies a length scale for the new 
dynamics.  We shall assume this scale to be of the order of kiloparsecs.

Note that Eq. (\ref{pares_eq_h00}) can be cast as a Poisson equation
for  
$\psi \equiv \frac{c^2}{2} (h_{00} + \frac{\delta \phi}{\langle \phi \rangle})$

\begin{equation} 
\nabla^2 \psi = \frac{G{_N}}{1+\alpha} \, 4\pi \rho \; ,
\label{pares_eq_psi}
\end{equation}
thus, the new Newtonian potential is now given by
\begin{equation} \label{phi-new}
\Phi_N \equiv \frac{c^2}{2} h_{00} = \psi - \frac{c^2}{2}
\frac{\delta \phi}{\langle \phi \rangle} \, .
\end{equation}

Solutions to these equations, the so--called potential--density
pairs \cite{BiTr08}, were  found for the NFW's and Dehnen's
density profiles \cite{RoCe04} and for axisymmetric systems
\cite{RoCePeTlCa05}.  For point particles the solution is well
known, see for instance \cite{He91,RoCe04}, and with the choice of
the above--mentioned background field, one has:
\begin{equation} \label{phi_New} 
\Phi_N = -\frac{G_{N}}{1+\alpha} \frac{M}{r} (1+ \alpha
e^{-r/\lambda}) \, ,
\end{equation}
where $M$ is the point particle mass producing the field.  The 
strength of the new scalar force is given
by $\alpha$ and its action range by $\lambda$. For local
scales, $r \ll \lambda$, deviations from the Newtonian theory are
exponentially suppressed, and for $r \gg \lambda$ the Newtonian
constant diminishes (augments) to $G_{N}/(1+\alpha)$ for positive
(negative) $\alpha$. Recently, the effect
of STT has been investigated in different cosmological scenarios in
which variations of the Newtonian constant are constrained.  For
instance, \cite{UmIcYa05} studied the influence of varying $G_N$ on
the Doppler peaks of the CMBR, and concluded that their parameter
($\xi= G/G_N$) should be in the interval $0.75 \le \xi \le 1.74$ in
order to be within the error bars of the CMBR measurements. In our
notation this translates into $-0.43 \le \alpha \le 0.33$. However,
this range for $\alpha$ has to be taken as a rough estimation, since
these authors have only considered  a variation of $G_N$, and not a
full perturbation study within STT. The latter has been done by
\cite{NaChSu02}, who found some deviations from the Newtonian
dynamics, that when translated into our strength parameter would
correspond to $\alpha = 0.04$; however, they do not compare their
results with observations. On the other hand, a structure formation
analysis has been done by \cite{ShShYoSu05}, in which deviations of
the matter power spectrum are studied by adding a Yukawa potential
to the Newtonian. They found some allowed dynamics, that turn out to
constrain our parameter to be within $-1.0 \le \alpha \le 0.5$; but
again a self-consistent perturbation study in general STT is
missing. Thus, the above three estimates can be taken as
order-of-magnitude constraints for our models. For definitiveness,
we will take values within the range $-0.3 \le \alpha \le 3$. The
value $\alpha=-0.3$ yields an asymptotic growing factor of $1.4$ in
$G_N$, whereas the value $\alpha=3.0$ makes $G_N$ to asymptotically 
reduce by one-fourth.

\section{A galactic model} \label{galactic-model}
We proceed to build a  galactic model by assuming that the
total matter content consists of two components, baryons and cold DM,
$\rho_{T}  = \rho_{\rm B} + \rho_{\rm DM}$;  Baryons represent stars, and
the cold DM component could be of any type. The dynamics is determined 
by the theory explained in the preceding section. There are two
possibilities on the DM origin:  i) DM is not related to the SF, and ii) DM is 
associated with the boson produced by the SF. In the former case, DM 
can be for example an ensemble
of neutralinos, whose mass is in the range $200 {\rm GeV} < m_\nu <
300 {\rm GeV}$ \cite{CaHeSu04}, within an effective supergravity
theory that nonminimally couples to gravity, see for example
\cite{El05}. In the latter case, ii), the mass of the DM particle is
given by $M_{DM} = h /\lambda c$, with $\lambda \sim$ kpc,
implying that $M_{DM} \sim 10^{-26}$eV.  The smallness of this mass
would have to be explained by a particle physics theory, e.g.
similar to light scalar presented by \cite{Ca98}, yet nonminimally
coupled to gravity.   

Both baryons and DM  `feel'  the same gravitational potential, 
$\Phi_N$, but are differently distributed in the galaxy. For the
baryon component we assume a  Freeman--disk density profile \cite{Fr70, BaSo80}, that is,
\begin{equation} \label{rho-baryons1}
\rho_B (r) = \frac{M_d}{2 \pi r_{d}^{2}}  \, e^{-r/r_{d}}
\end{equation}
where $M_d$ is the mass of the disk and  $r_d$ its radius. 

For the DM density  we do not assume a particular profile.
Instead, we proceed to find its form by imposing a general rotation
pattern.  In the past attempts have been made to determine a universal rotation  curve (URC) profile,
beginning with the pioneering work of Ref. \cite{RuBuFoTn85}. In the nineties the authors of 
Refs. \cite{PeSa91,PeSaSt96}  considered more than 
1100 optical and radio data of Sb-Im spirals to find a phenomenological  
URC profile valid out the outermost radius where 
data were  available at that time.  Recently, they have considered more data and have 
modified the profile  \cite{Sa07},  extending it out  to its virial radius, that is, including the DM 
halo part.  This profile is supposed to be valid for spirals of different types \cite{SaPe97} but a number of issues 
are still open \cite{Sa07}.  Accordingly, we assume for our model that  stars and DM particles obey the following 
URC profile \cite{Sa07}: 
\begin{widetext}
\begin{eqnarray} \label{ph-rot-prof} 
v_{URC}^{2} &=& v_{URCD}^{2} + v_{URCH}^{2}  \nonumber  \\   
&=&  \frac{G M_d }{2 r_{d}} \left(\frac{r}{ r_{d}}\right)^{2} 
\left[ {\rm I_{0}}\left(\frac{r}{2 r_{d}}\right) {\rm K_{0}} \left(\frac{r}{2 r_{d}}\right) -  
{\rm I_{1}}\left(\frac{r}{2 r_{d}}\right) {\rm K_{1}} \left(\frac{r}{2 r_{d}}\right) \right] \nonumber \\
&& + \frac{2 \pi G \rho_{0} r_{0}^{3}}{r} \Big\{ {\rm \ln}\left(1+ \frac{r}{r_{0}}\right) +  
\frac{1}{2}{\rm \ln}\left[1+ \left(\frac{r}{r_{0}}\right)^2 \right] - {\rm \arctan} 
\left(\frac{r}{r_{0}}\right) \Big\} = r \frac{d\Phi_N}{dr} , 
\end{eqnarray}
\end{widetext}
where the functions I and K are the modified Bessel functions, and  $\rho_{0}$ and $r_{0}$ are the scaling density 
and radius  of the Burkert density profile \cite{Bu95}. The first part accounts for the disk contribution and the second for the halo's.
In a previous work we have assumed  a simpler, flat rotation curve profile \cite{CeRoNu07}.

The given circular velocity determines the form of gravitational potential $\Phi_N$, through Eq. (\ref{ph-rot-prof}), which in turn 
is related to $\rho_{dm}$ and $\delta \phi$ through Eqs. (\ref{pares_eq_h00},\ref{pares_eq_phibar},\ref{phi-new}). 

\begin{figure*}
\includegraphics[width=88mm]{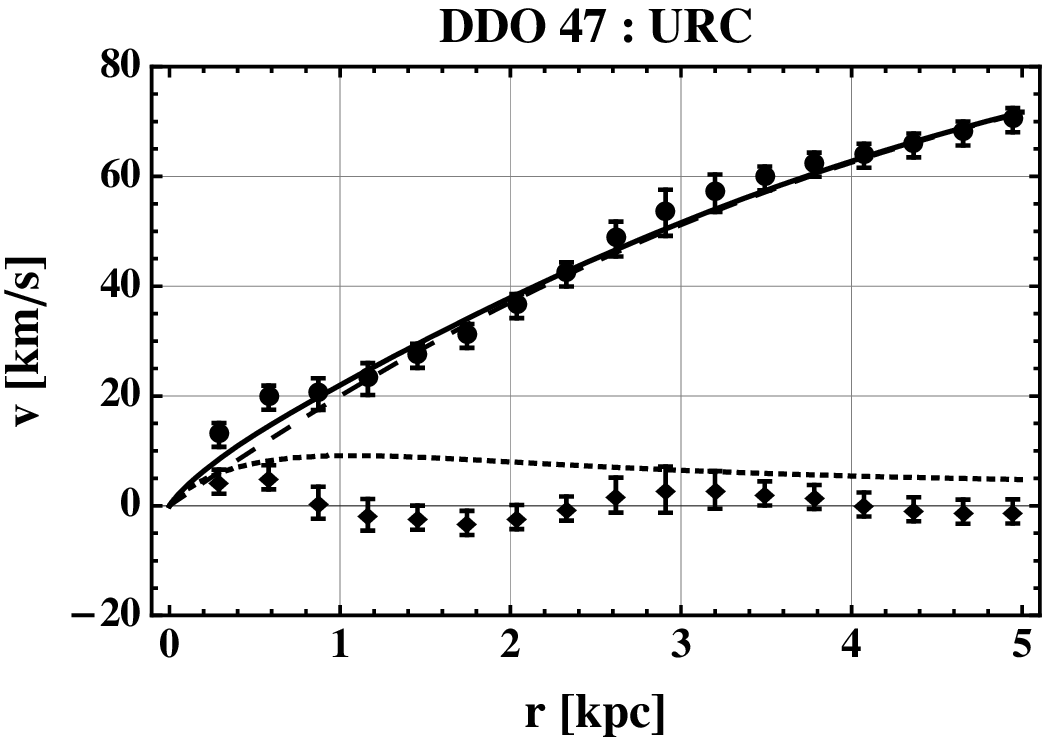} 
\includegraphics[width=88mm]{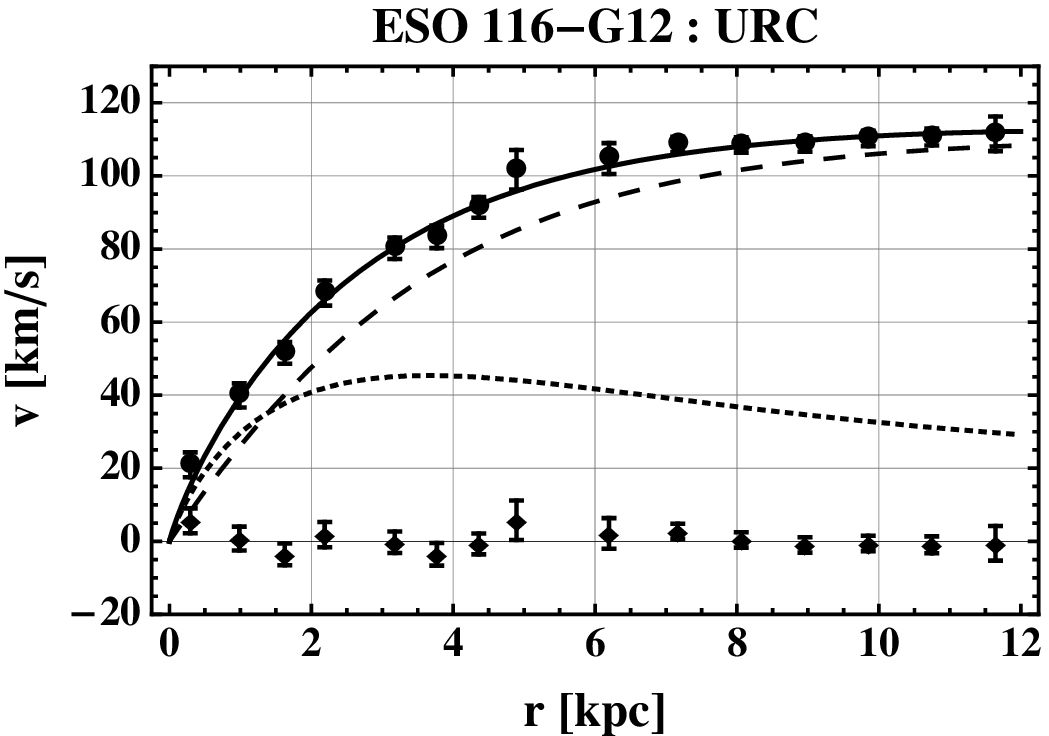} 
\includegraphics[width=88mm]{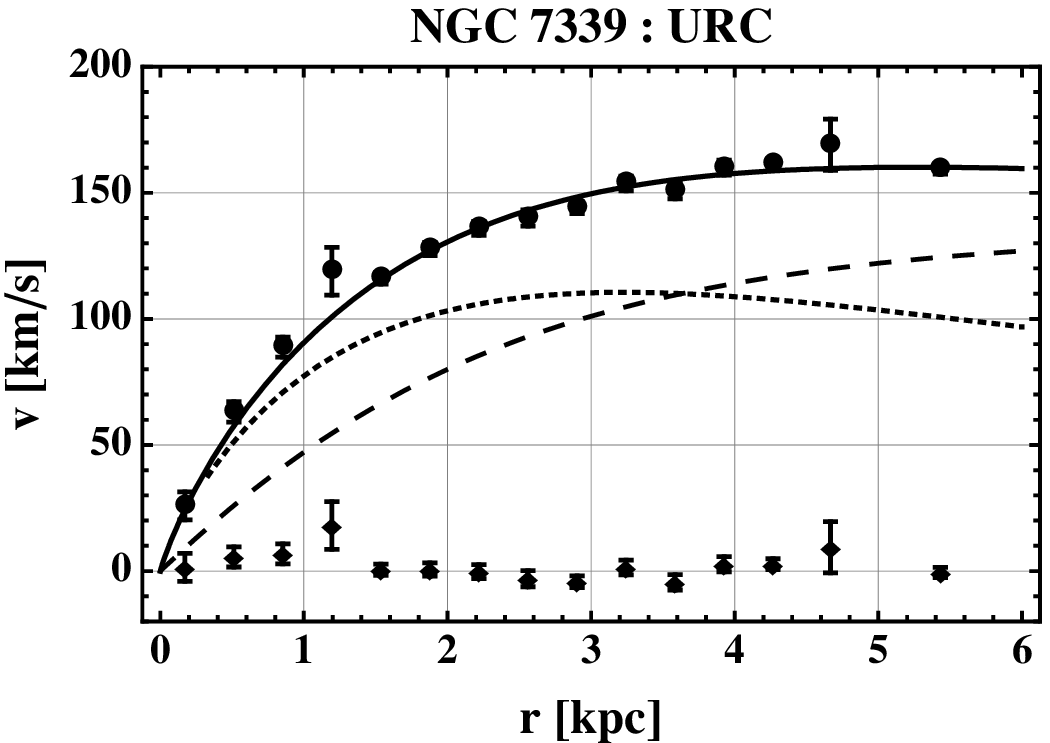} 
\includegraphics[width=88mm]{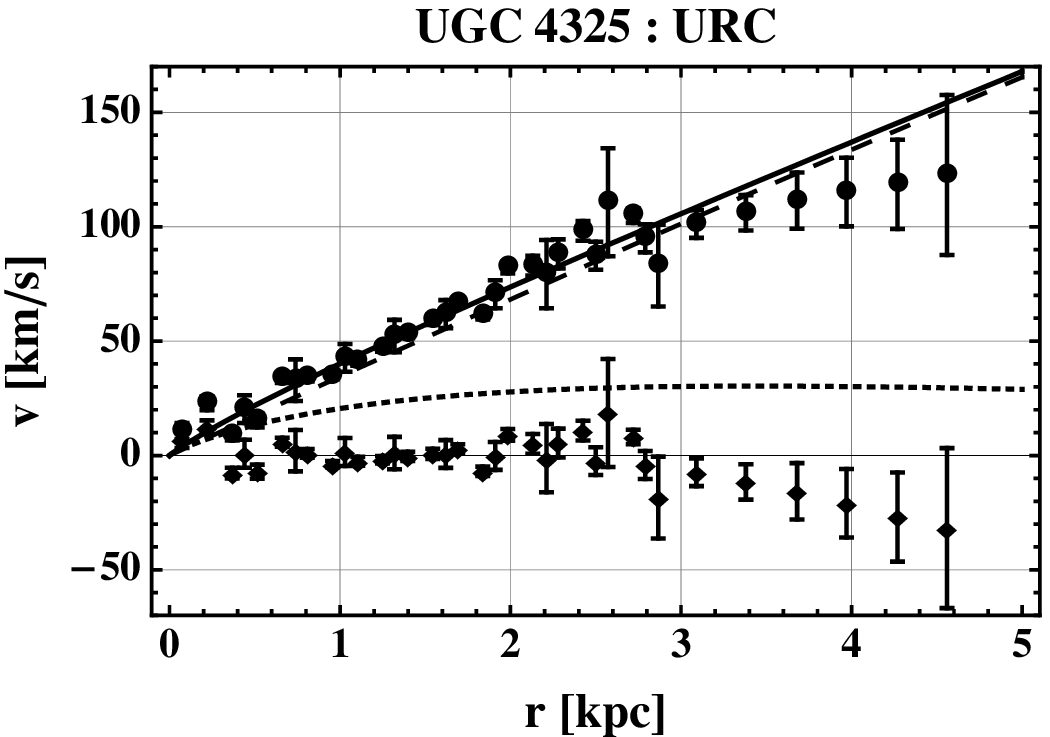} 
\caption{The continuous line is the fitted rotation curve for data of the four galaxies.  The short-dashed line is the exponential disk contribution and  
the long-dashed line is DM's.  At the bottom we plotted the residuals ($v_{\rm obs}-v_{URC}$).} 
\label{rcs}
\end{figure*}

Integrating Eq. (\ref{ph-rot-prof}) for $\Phi_N$ yields
\begin{widetext}
\begin{eqnarray} \label{ph-N}  
\Phi_N &=& \Phi_{ND} + \Phi_{NH}  \nonumber \\
&=&   - \frac{1}{2}  \frac{G M_{d}}{r_d}  \frac{r}{r_d}  \left[ {\rm I_{0}}\left(\frac{r}{2 r_{d}}\right) {\rm K_{1}} \left(\frac{r}{2 r_{d}}\right) -  
{\rm I_{1}}\left(\frac{r}{2 r_{d}}\right) {\rm K_{0}} \left(\frac{r}{2 r_{d}}\right) \right] \nonumber \\
&& 
+\frac{2 \pi G \rho_{0} r_{0}^{3}}{r}   \Big\{ \left(1+ \frac{r}{r_{0}}\right) \left[ \arctan \left(\frac{r}{r_{0}} \right)  
- \ln \left(1+ \frac{r}{r_{0}} \right) \right]  + \frac{1}{2}  \left(-1+ \frac{r}{r_{0}}\right) \ln \left(1+ \frac{r^2}{r_{0}^2} \right)  
\Big\} 
\, , 
\end{eqnarray}
which leads to the motion of test particles in the Galaxy.  Substituting this result in 
the original system, Eqs. (\ref{pares_eq_h00}, \ref{pares_eq_phibar}) transform into 
the following two equations
\begin{eqnarray} 
\nabla^2 \delta \phi - \frac{m^2}{(1+\alpha)} \, \delta \phi & = & - 
\frac{2 \alpha}{G_N} \, \nabla^2 \Phi_N \; ,
\label{phibar} \\ 
 \rho_{DM} &= & -\rho_B + \frac{1}{ 4\pi G_{N}} \nabla^2 \Phi_N +
\frac{1}{8 \pi (1+\alpha)} \left(\frac{m c}{h}\right)^{2} \delta \phi   =  \frac{\rho_0 \, r_{0}^{3}}{(r+r_{0}) (r^{2}+r_{0}^{2})} +
\frac{1}{8 \pi (1+\alpha)} \left(\frac{m c}{h}\right)^{2} \delta \phi   \; , \qquad
\label{rho-dm}
\end{eqnarray}
\end{widetext}
for two variables, $\rho_{DM}$ and $\delta \phi$. The second equality of Eq. (\ref{rho-dm}) results since the disk 
contribution of Laplacian of $\Phi_N$ cancels out with baryon density, see Eq. (\ref{ph-N}).   Thus, the  resulting DM profile is 
the Burkert profile plus the SF  contribution.  
By substituting Eq. (\ref{ph-N}) into Eq. (\ref{phibar}), one solves for
the scalar perturbation. Then, using Eq. (\ref{rho-dm}) one solves for the DM profile.   The results depend on the STT 
parameters $(\lambda, \, \alpha)$ and on the rotation curve fitting parameters ($ \rho_{0}, r_{0}, M_{d}, r_{d} $).   Particular 
galaxies fix the latter parameters. As an example, we had chosen a set of four galaxies of different types that have been 
used  to test particular gravity theories \cite{FrSa07} and to test the validity of some density  
profiles \cite{Ge04,KuMcBl08}. The galaxies are:  DD047, ESO 116-G12, NGC 7339 and UGC 4325.   A greater galaxy set   
could have been used to test the rotation profile given by Eq. (\ref{ph-rot-prof}),  but this has already been done in Ref. \cite{Ge04},  
and in Ref. \cite{Sa07} additional arguments are given in favor of this rotation profile.  In general, flat density profiles, such as Burkert's, tend to 
match rotation curves data within Newtonian dynamics \cite{Ge04}.   In table \ref{table1} we present the 
properties and  best fitted values of the galaxies' parameters, and in Fig. \ref{rcs} the best fits are shown. The rotation curve data 
fit quite well to the profile URC given by Eq. ({\ref{ph-rot-prof}}).  In doing the fitting we have taken the values for  $r_d$, which is directly 
measured by optical observations,   given in Ref. \cite{FrSa07b} for the first three galaxies and the values given in Ref. \cite{KuMcBl08} for 
UGC 4325.  Then,  we have varied $M_d$, $r_0$ and $\rho_0$.

%
%
{\renewcommand{\arraystretch}{2.5}\begin{table*}
\caption[Galaxy properties and parameters]{Properties and best fitting parameters of the galaxies used. \label{table1}}
\begin{displaymath}
\begin{array}{ccccccc}\hline\hline
{\rm Galaxy } & \quad\quad  {\rm Type} \quad\quad  &r_{d} \, [{\rm kpc}] & {M_{d}} \, 
[M_{\odot}]& r_{0} \, [{\rm kpc}]& \rho_{0} \, [M_{\odot}/{\rm kpc}^3 ]& 
\chi^{2}_{\rm red}\\ \hline  \hline  
{\rm DDO}\, 47 & {\rm IB}& 0.5 & 3.60 \pm 0.62 \times 10^{7} &5.43 \pm 0.09& 2.67 \pm 0.03  \times 10^{7}   & 1.74\\ \hline  
{\rm ESO} \,116-G12 & {\rm SBcd}& 1.7  & 2.09 \pm 0.08 \times 10^{9} &4.77 \pm 0.03 & 4.44 \pm 0.04 \times 10^{7}  & 0.99\\ \hline  
{\rm NGC}\, 7339 & {\rm SABd}& 1.5  & 1.10 \pm 0.01 \times 10^{10} &3.03\pm 0.03 & 1.60 \pm 0.02 \times 10^{8}   & 1.69\\ \hline  
{\rm UGC}\, 4325 & {\rm SA}& 1.6  & 8.42 \pm 0.47 \times 10^{8} &40.54 \pm 10.16 & 6.59 \pm 0.08 \times 10^{7}  & 3.56\\ \hline \hline 
\nonumber\end{array}
\end{displaymath}\end{table*}}

The first term of the right-hand-side of Eq. (\ref{rho-dm}) is the Burkert profile, 
$\rho_{BUR} \equiv \rho_0 \, r_{0}^{3}/(r+r_{0}) (r^{2}+r_{0}^{2})$ and one can define 
$ \rho_{\phi} \equiv \frac{1}{8 \pi (1+\alpha)} \left(\frac{m c}{h}\right)^{2} \delta \phi$ so that to 
express  Eq. (\ref{rho-dm}) as 
$\rho_{DM}  =   \rho_{BUR} +    \rho_{\phi}$.   The density $ \rho_{\phi}$ is the contribution of the SF fluctuation to the 
total density. Given this, the total density is 
$\rho_{T}= \rho_{B} + \rho_{DM}   =    \rho_{B} + \rho_{BUR} +    \rho_{\phi} $.   In what follows  we 
proceed to numerically integrate the above equations.  

\section{Spherical solution} \label{spherical-sol}

Unfortunately, these quantities cannot be computed analytically, since the Newtonian potential involves 
cylindrical and spherical dependencies.  Therefore to solve the Eq. (\ref{phibar})  we will assume 
spherical symmetry.  Then, we use the integrals given in Ref.  \cite{RoCe04} to 
numerically solve for $\delta \phi$. 

We perform the integration for the LSB galaxy UGC 4325;  because the NFW model does not properly fit with  the observations, so we plan for this galaxy to contrast  
our results with them. 

Let us explain the range of values for the STT parameters taken in our analysis.  Originally,  STT \cite{BrDi61,Wa70} were 
thought for positives values of $\alpha$ to have a standard kinetic term in Eq. (\ref{SFLagrangian}).  But negatives values are also theoretically 
possible \cite{Ni84,Po88} and they have been applied, for instance, to accomplish, without a 
potential in Eq. (\ref{SFLagrangian}),  an inflationary era  in  isotropic \cite{Le95} and anisotropic models \cite{Ce99}, or more recently, to 
explain the present accelerated expansion of the Universe in some quintessence models \cite{Ca02,Da03}.   Thus, we will consider 
positive as well negative values  of $\alpha$ subject to the constraints 
mentioned at the end of section \ref{newapprox}.  Therefore, we will analyze the solutions in the interval $-0.3 \le \alpha \le 3.0$. On 
the other hand, we will assume that $\lambda$ is in the interval $0.1 \,{\rm kpc} \le \lambda \le 50$ kpc to fit galactic scales.

\subsection{Solutions for positive  $\alpha$}

\begin{figure*}
\includegraphics[width=150mm]{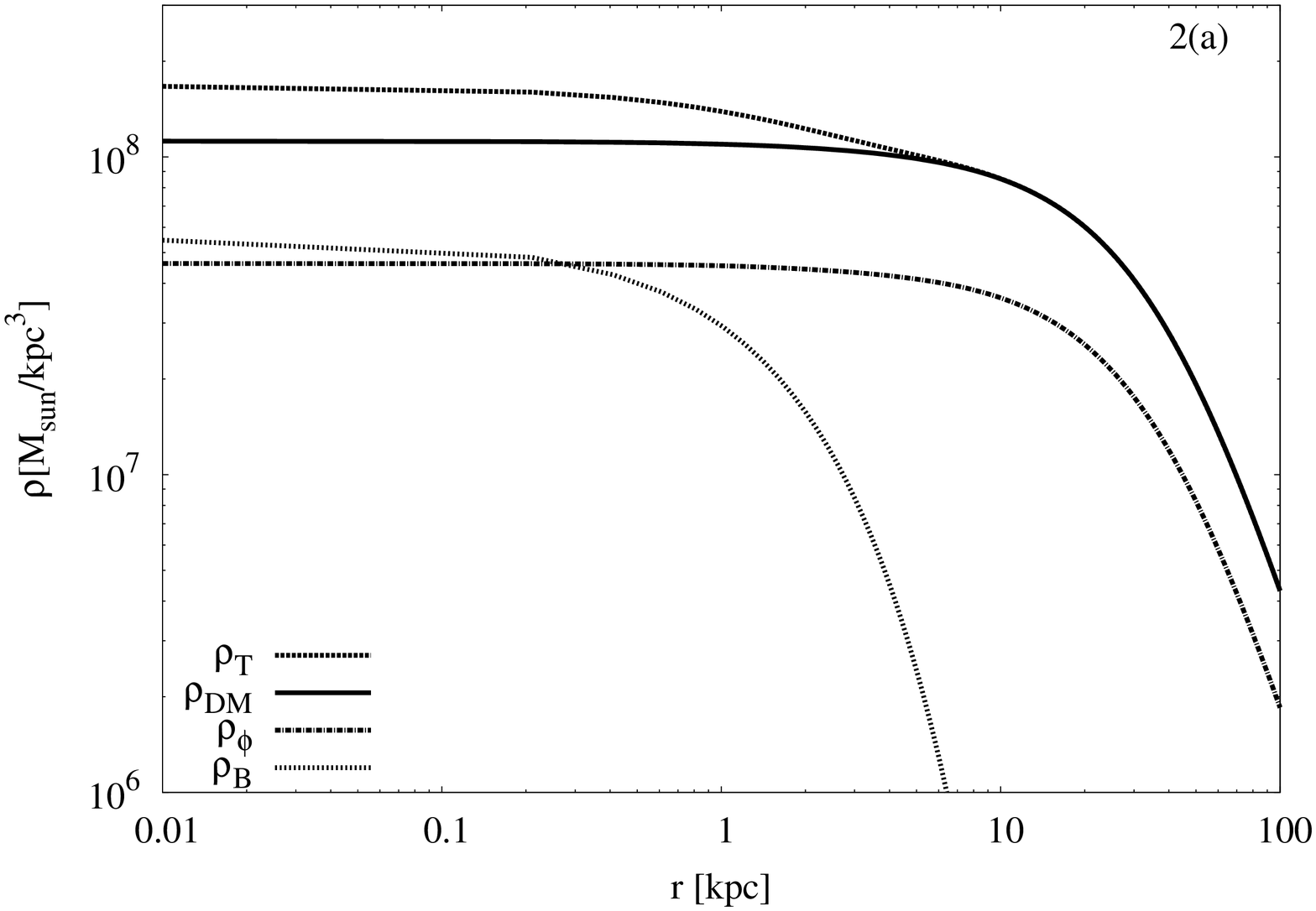} 
\includegraphics[width=150mm]{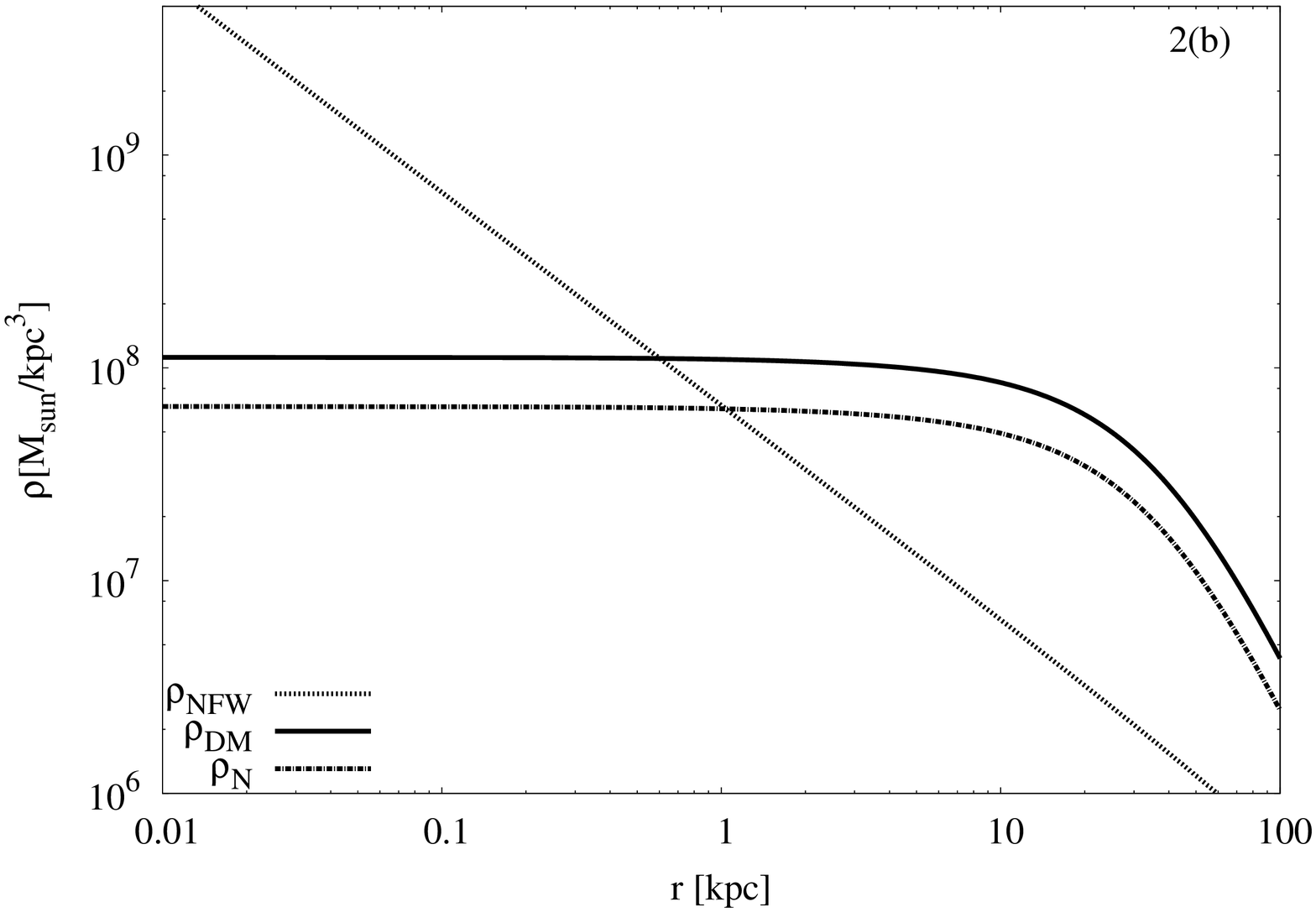} 
\caption{ {\textbf (a)} It is shown all density profiles using $\alpha=3.0$ and $\lambda=1$ kpc. {\textbf (b)} It is 
shown $\rho_{NFW}$,  $\rho_{DM}$ and $\rho_{N}$ for comparison.} 
\label{all-densities}
\end{figure*}

In Fig. \ref{all-densities}a we plot the resulting density profiles for $\alpha=3.0$ and $\lambda=1.0$ kpc.  Of particular 
interest is the form of the DM profile that flattens near the center of the Galaxy.  The latter is 
again shown in Fig. \ref{all-densities}b, where for comparison  the {\textit standard} Newtonian 
density profile, $\rho_{N}$, is plotted, which is the profile to have  $v^{2}_{URCH}$ as given by Eq. (\ref{ph-rot-prof}), but turning off 
the SF; i.e., $\rho_{N}$ is the density found by solving the Poisson equation with Newtonian potential given by $\Phi_{NH}$ in   
Eq. (\ref{ph-N}).  Thus,  the Newtonian density is the Burkert profile \cite{Bu95}.   One observes that  $\rho_{DM} $ is always bigger than $ \rho_{N} $,  since the effect of the SF is to diminish the 
effective gravitational constant for $r>\lambda$, since   
$G_{\rm eff}= G (1+\alpha e^{-r/\lambda})/(1+ \alpha) $, thus to compensate the reduction 
of the gravitational constant, a denser DM profile is necessary to have the same rotation 
curve profile. Concerning the behavior of the profile, we have computed a numerical fit of the inner part
of the curve ($r \ll r_{d}$), showing that it behaves approximately as
$\rho_{\rm DM} \sim r^{-\gamma_{DM}}$ with $\gamma_{DM} \approx 0.00006^{+0.00002}_{-0.00001} $; the uncertainties stemming 
basically from the uncertainties in $r_0$.  On the other hand,  the 
standard Newtonian model is just a little steeper 
$\rho_{\rm N} \sim r^{-\gamma_{N}}$ with $\gamma_{N} \approx 0.00010^{+0.00003}_{-0.00002}$. Both of these profiles are 
essentially shallow.  The  NFW  that best fitted rotation curves data is included for comparison, which is known to be cuspier in the inner 
region,  $\rho_{\rm NFW} \sim r^{-1.00001}$.  For $r \sim r_{d}$  the behavior follows  
$\rho_{\rm DM} \sim r^{-\delta_{DM}}$ with $\delta_{DM} \approx  0.040^{+0.011}_{-0.007}$,  
 $\rho_{\rm N} \sim r^{-\delta_{N}}$ with $\delta_{N} \approx 0.041^{+0.014}_{-0.008}$. The NFW model 
 behaves as  $\rho_{\rm NFW} \sim r^{-\delta_{NFW}}$ with  $\delta_{NFW} \sim 1.003$. 
 For $r \sim \, r_{0}$  the behavior follows  $\rho_{\rm DM} \sim r^{-\delta_{DM}}$ with $\delta_{DM} \approx  1.44^{+0.34}_{-0.26}$,  
 $\rho_{\rm N} \sim r^{-\delta_{N}}$ with $\delta_{N} \approx 1.44^{+0.34}_{-0.26}$, and the NFW model 
 behaves as  $\rho_{\rm NFW} \sim r^{-\delta_{NFW}}$ with $\delta_{NFW} \sim 1.08$.   Fig. \ref{all-densities}b shows that the 
 DM profile is bigger than NFW's beyond $r \sim 0.6$ kpc.  On the other hand,  Ref. \cite{KuMcBl08} shows that NFW fits are bad for this galaxy, because  
 arbitrary low concentrations (i.e. large NFW scaling length $r_s$) are needed.  In our case, the best fitted curve implies a concentration $c=1.22$, which is  
 clearly inconsistent with the cosmological expected values $c >5$ (for a wide range of $v_{200}$ values) \cite{Te04,Sp07}.  Thus, what 
 we essentially see in our plots of the NFW profile is its cuspy region, as shown in Fig.  \ref{all-densities}b.  
 
 The solution shown for the SF is the interior 
solution and eventually at some $r=R$ the exterior solution is valid \cite{RoCe04}; one could think of $R$ 
to be of the order of the halo size. Thus for $r\ge R$, the SF exponentially vanishes and we obtain a 
standard Newtonian behavior.   Therefore,  asymptotically, for $r \gg r_{0}$, $\rho_{\rm DM} \sim r^{-3}$ similar to  
 $\rho_{\rm NFW} \sim r^{-3}$. 

\begin{figure*}
\includegraphics[width=150mm]{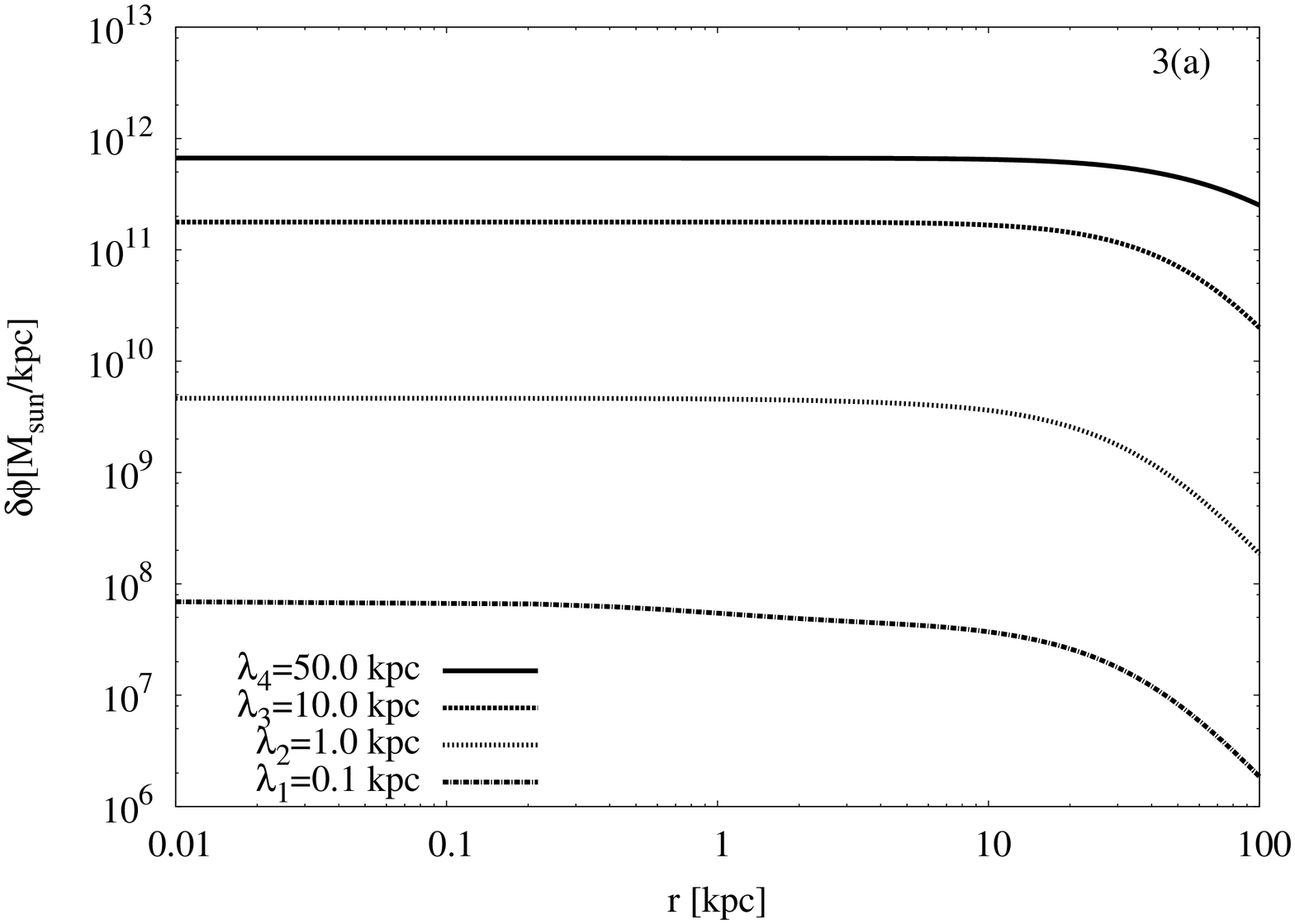}
\includegraphics[width=150mm]{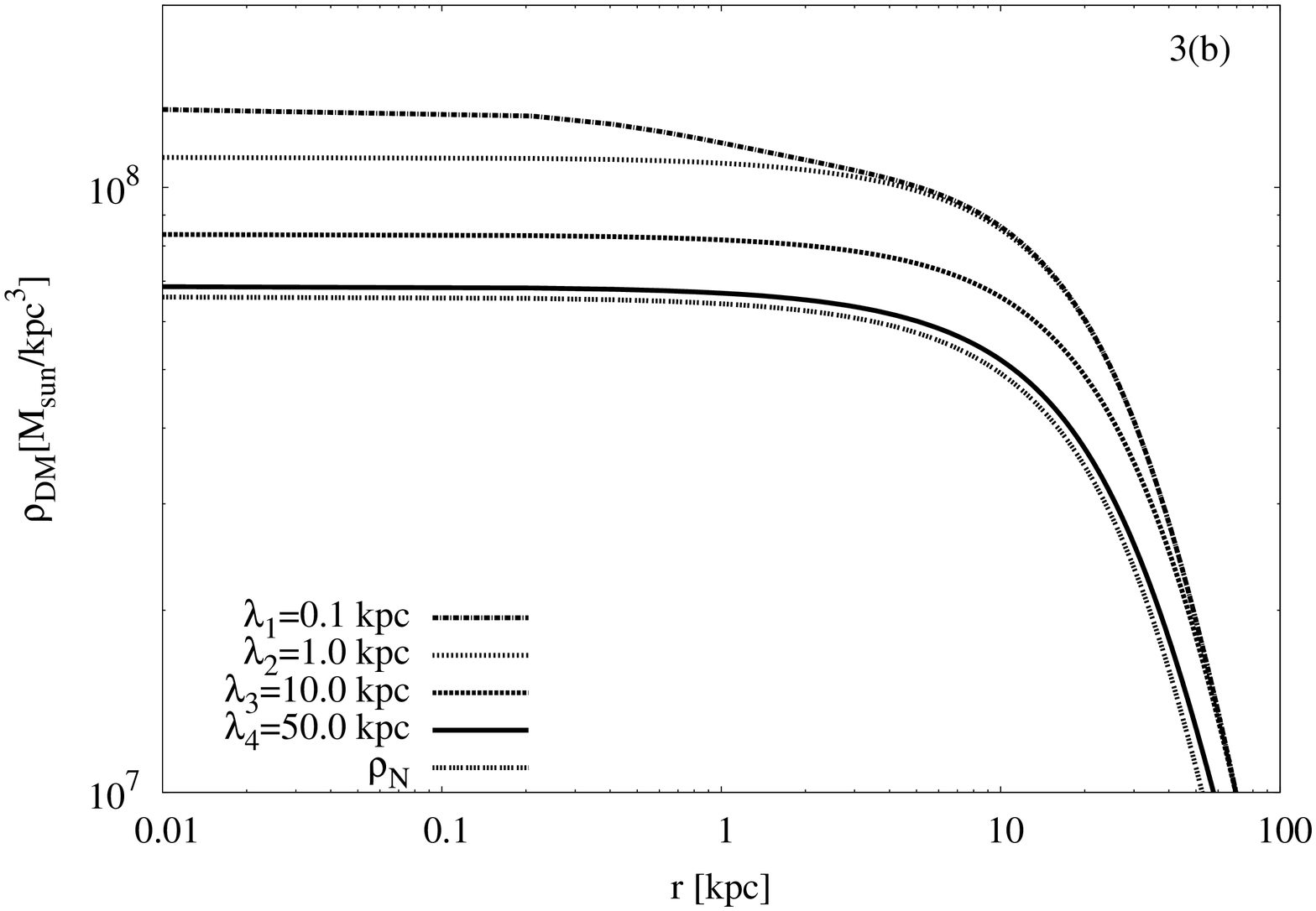} 
\caption{{\textbf (a)} It is shown the SF perturbation and {\textbf (b)} the DM profile for various $\lambda$ and fixed $\alpha=3.0$.}
\label{phi-alpha-lambdas}
\end{figure*}

In Figs.  \ref{phi-alpha-lambdas}a-b we have plotted the SF and 
DM profiles for various $\lambda$.  The SF fluctuation diminishes 
going from the center to outer parts, and for smaller $\lambda$ the decay is stronger  in inner regions.  Fig. 
\ref{phi-alpha-lambdas}a shows also that the SF  
is bigger for larger $\lambda$, since for large $\lambda$, $G_{\rm eff}$ approaches to $G $.    A consistency check implies that    
the SF must comply with the condition  
$\delta \phi <  (1+\alpha) c^{2} G_{N}^{-1}  =  (1+\alpha) \, 2.1 \times 10^{16} M_{\odot}/$kpc in order to validate the 
perturbation theory used, and it is indeed the case for the pair of parameters $(\lambda, \, \alpha)$ 
chosen.  On the other hand, the DM profiles in Fig. \ref{phi-alpha-lambdas}b  diminish by 
augmenting $\lambda$. The DM profile more deviates from the standard Newtonian one for smaller  $\lambda$.

\begin{figure*}
\includegraphics[width=150mm]{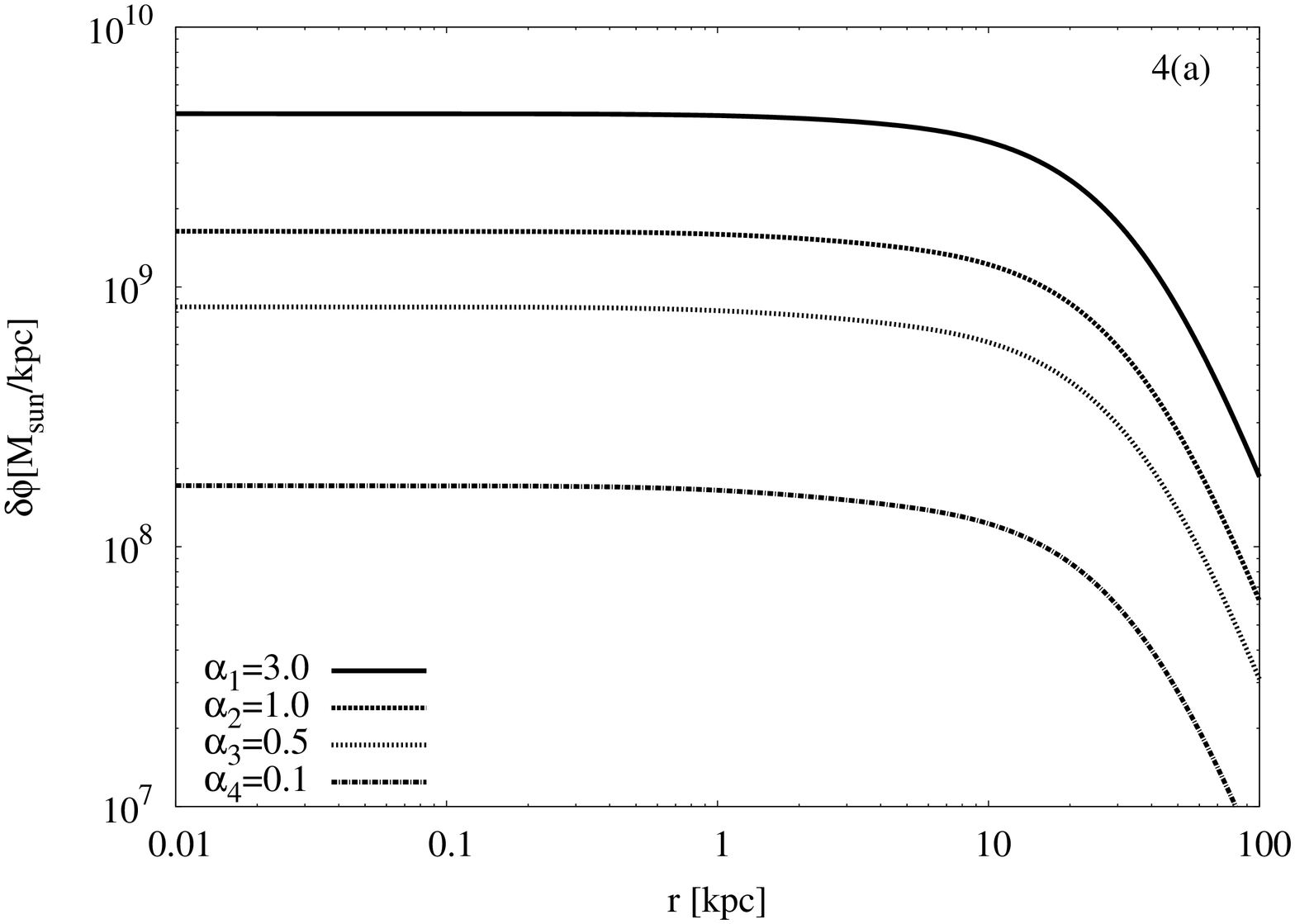}
\includegraphics[width=150mm]{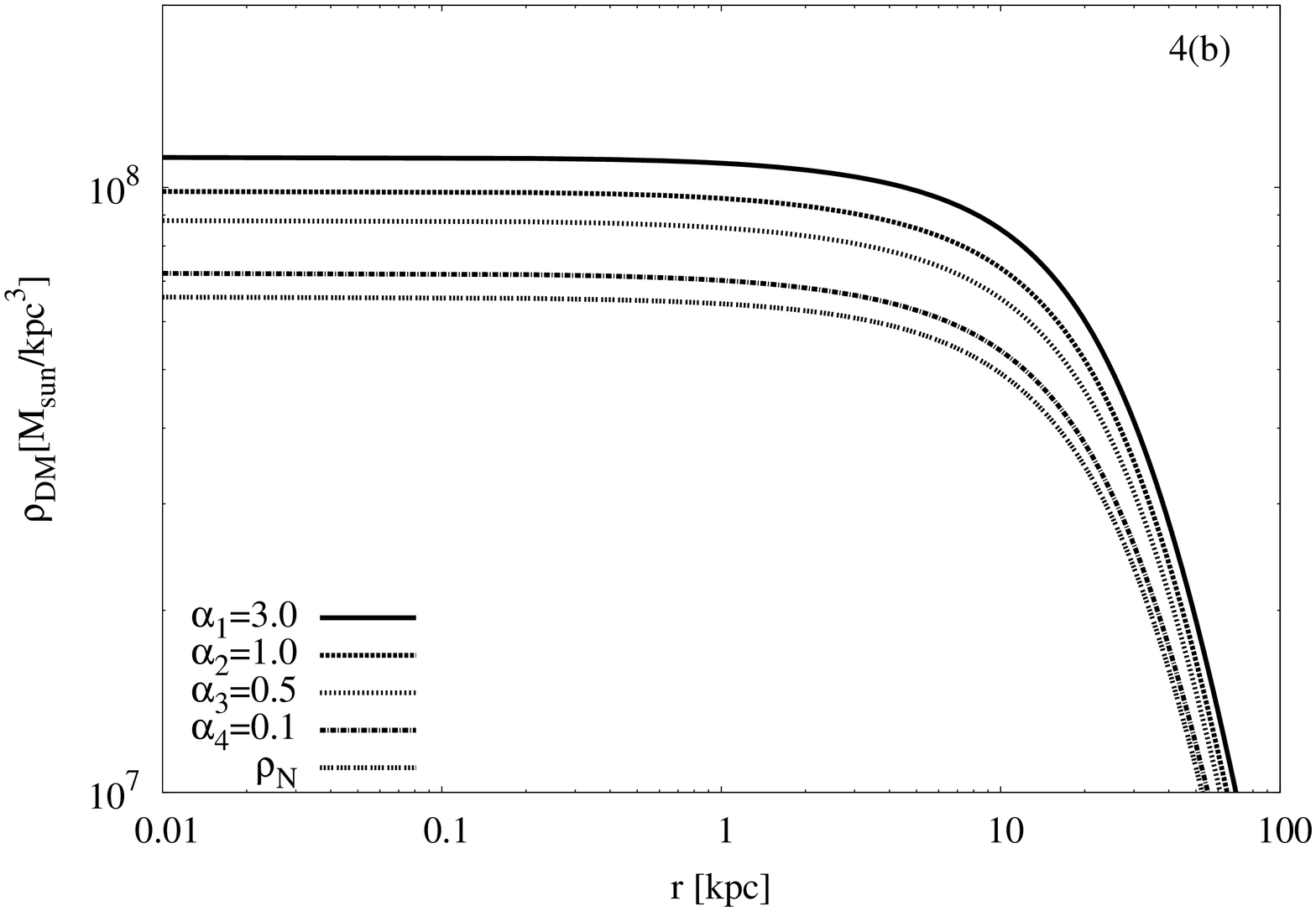} 
\caption{{\textbf (a)} It is shown the SF perturbation and {\textbf (b)} the DM profile for various $\alpha$ and fixed $\lambda=1$ kpc.}
\label{phi-lambda-alphas}
\end{figure*}

In Figs. \ref{phi-lambda-alphas}a-b we plotted the SF and DM profiles for  
various $\alpha$ and fixed $\lambda = 1.0$ kpc.  Again the constraint on  $\delta \phi $ is 
fulfilled (Fig. \ref{phi-lambda-alphas}a).  As expected, for small $\alpha$ the DM profile 
tends to the standard Newtonian one (Fig. \ref{phi-lambda-alphas}b). 

\subsection{Solutions for negative  $\alpha$}

\begin{figure*}
\includegraphics[width=150mm]{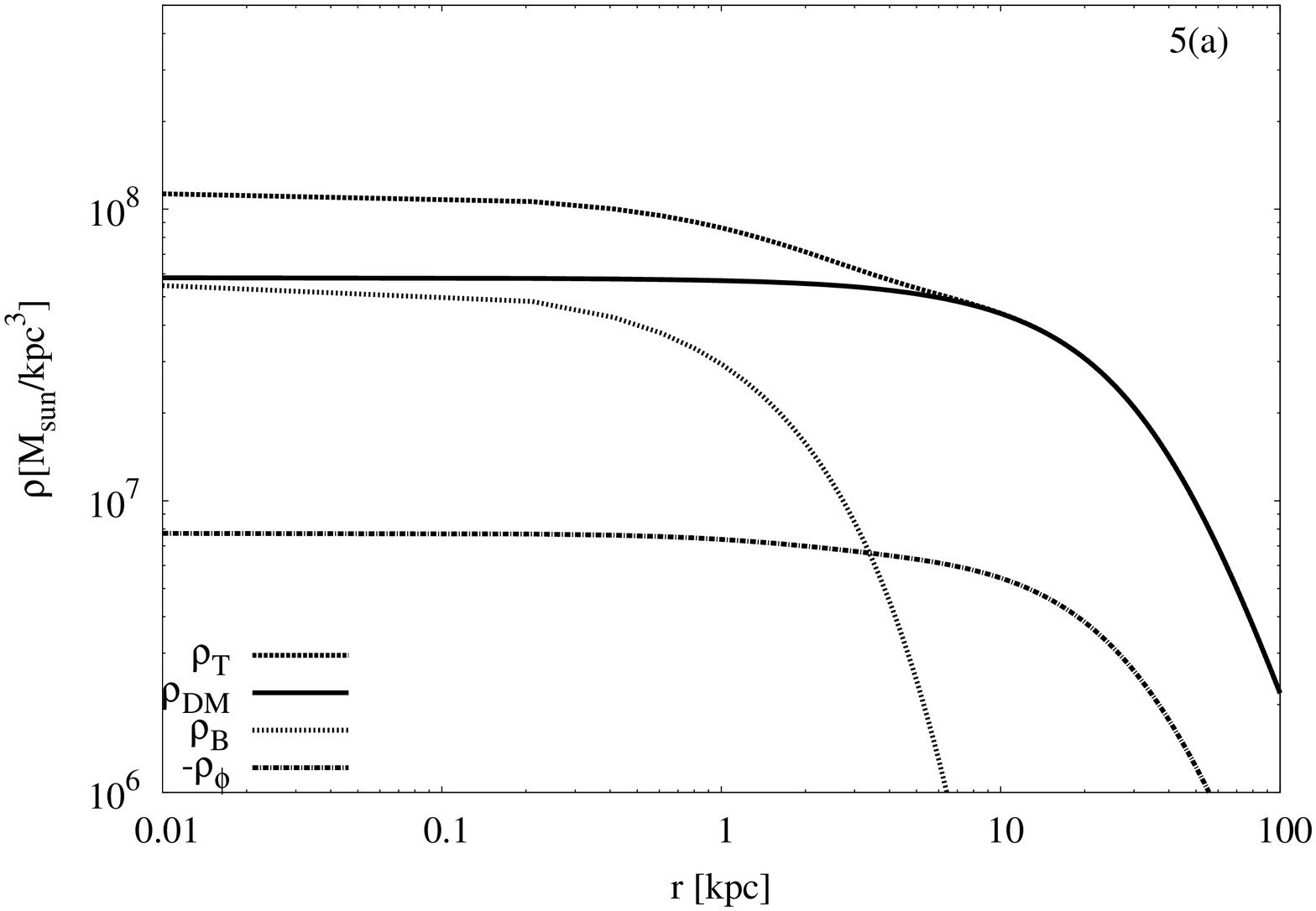} 
\includegraphics[width=150mm]{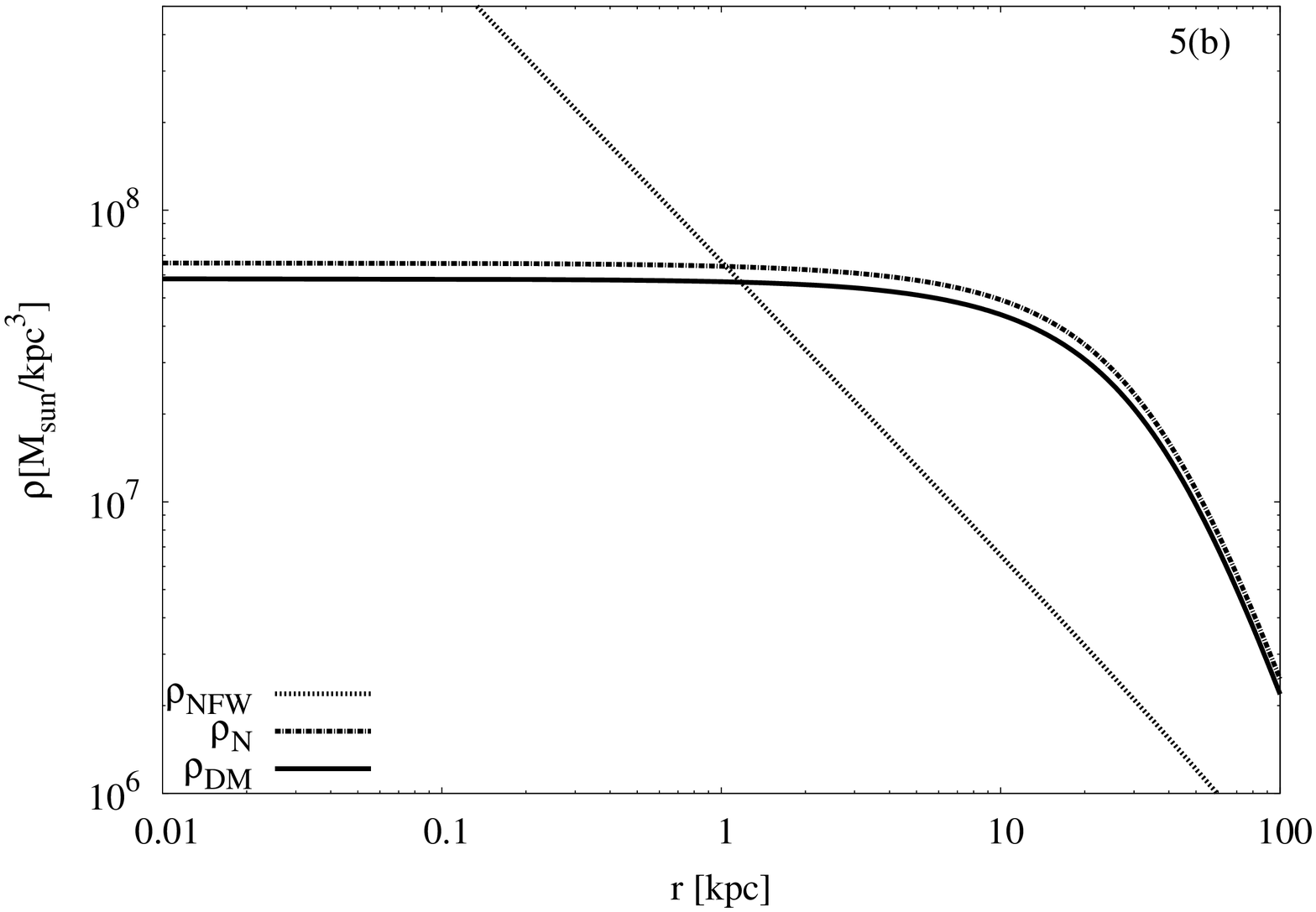}  
\caption{{\textbf (a)} It is shown all density profiles using  $\alpha=-0.1$ and $\lambda=1$ kpc. {\textbf (b)} 
It is shown $\rho_{NFW}$,  $\rho_{DM}$ and $\rho_{N}$ for comparison.}
\label{all-densities-n}
\end{figure*}

In Fig. \ref{all-densities-n}a we again plot all densities, as in Fig. \ref{all-densities}a, but 
now for $\alpha=-0.1$ and $\lambda=1.0$ kpc.  
The DM profile  in the inner regions ($r \ll r_{\rm d}$) is given by   
$\rho_{\rm DM} \sim r^{-\gamma_{DM}}$ with $\gamma_{DM} \approx 0.00011^{+0.00004}_{-0.00002}$, that is quite similar as the standard Newtonian, which gives 
$\rho_{\rm N} \sim r^{-\gamma_{N}}$, with $\gamma_{N} \approx 0.00010^{+0.00003}_{-0.00002}$;  they are shallow. In contrast, NFW's profile in the 
same region that has a cuspy power $\gamma_{NFW}  = -1.00001$.  Fig. \ref{all-densities-n}b shows the DM, Newtonian and NFW 
profiles for comparison. It is clear that the DM profile is less massive than the standard Newtonian. This 
is because the effective gravitational function, $G_{\rm eff}= G (1+\alpha e^{-r/\lambda})/(1+ \alpha) $, is 
increased for negative $\alpha$, and the gravitational pull, being proportional to the term $G_{\rm eff} M$, is 
compensated by a decrease in $M(r)$, that is by a decrease of the DM profile.  For $r \sim r_{d}$  the 
behavior is as follows: $\rho_{\rm DM} \sim r^{-\delta_{DM}}$ with  
$\delta_{DM} \approx  0.035^{+0.015}_{-0.009}$, the standard Newtonian  profile is    
$\rho_{\rm N} \sim r^{-\delta_{N}}$ with $\delta_{N} \approx 0.041^{+0.014}_{-0.008}$,  the NFW's  behaves 
nearly as $\delta_{NFW}= 1.00$.   For $r \sim  \,r_{0}$  the behavior follows  $\rho_{\rm DM} \sim r^{-\delta_{DM}}$ 
with $\delta_{DM} \approx  1.44^{+0.34}_{-0.26}$,   $\rho_{\rm N} \sim r^{-\delta_{N}}$ with $\delta_{N} \approx 1.44^{+0.34}_{-0.26}$, the NFW 
 model  behaves as  $\rho_{\rm NFW} \sim r^{-\delta_{NFW}}$ with $\delta_{NFW} \sim 1.08$.  As in the $\alpha$-positive 
 case, Fig.  \ref{all-densities-n}b shows again that beyond some region ($r \sim 1$ kpc) the DM profile is bigger than NFW's.  
Asymptotically, for $r \gg r_{0}$, beyond some point the exterior solution is valid, thus  $\rho_{\rm DM} \sim r^{-3}$,  
similar to  $\rho_{\rm NFW} \sim r^{-3}$.

\begin{figure*}
\includegraphics[width=150mm]{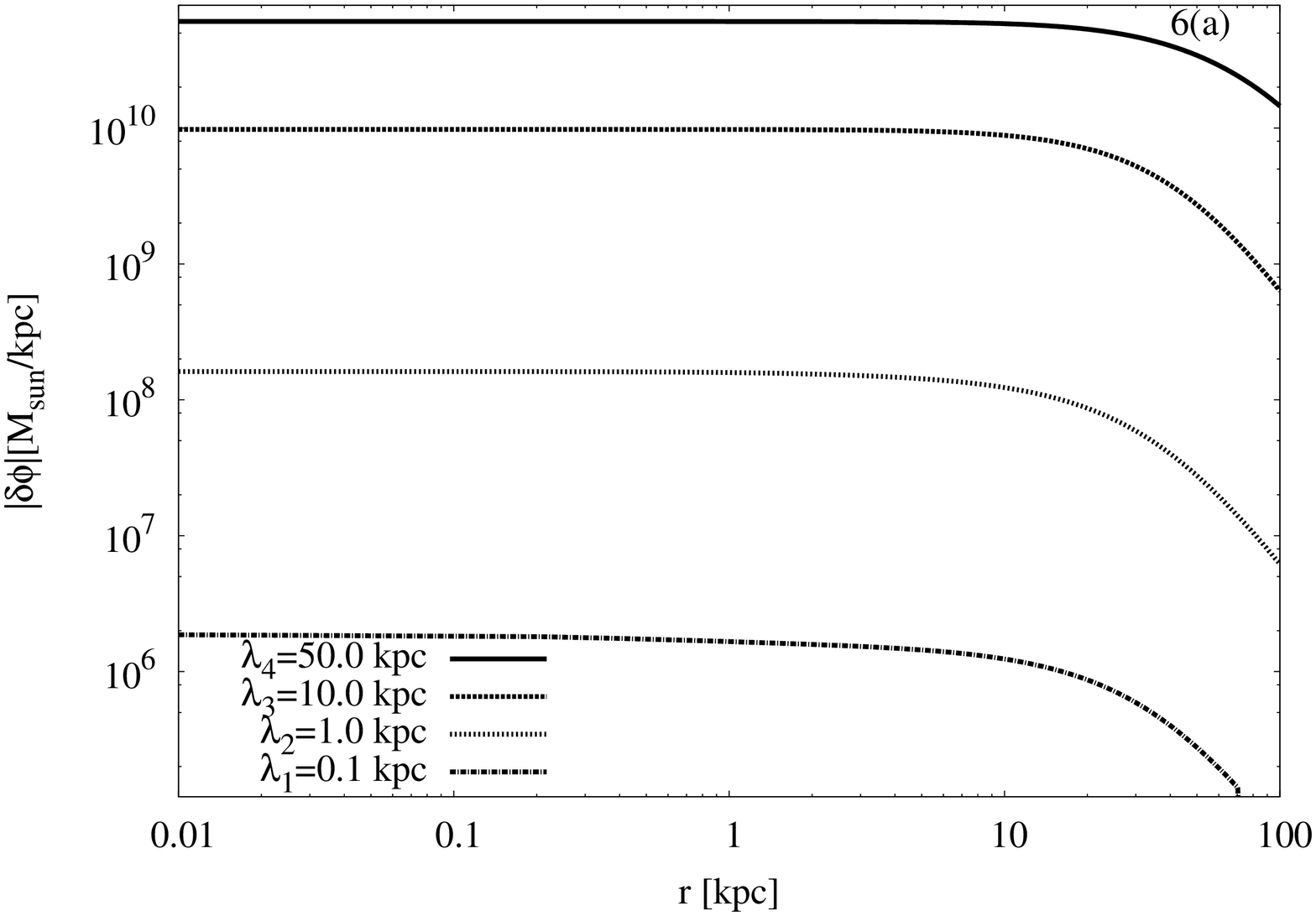} 
\includegraphics[width=150mm]{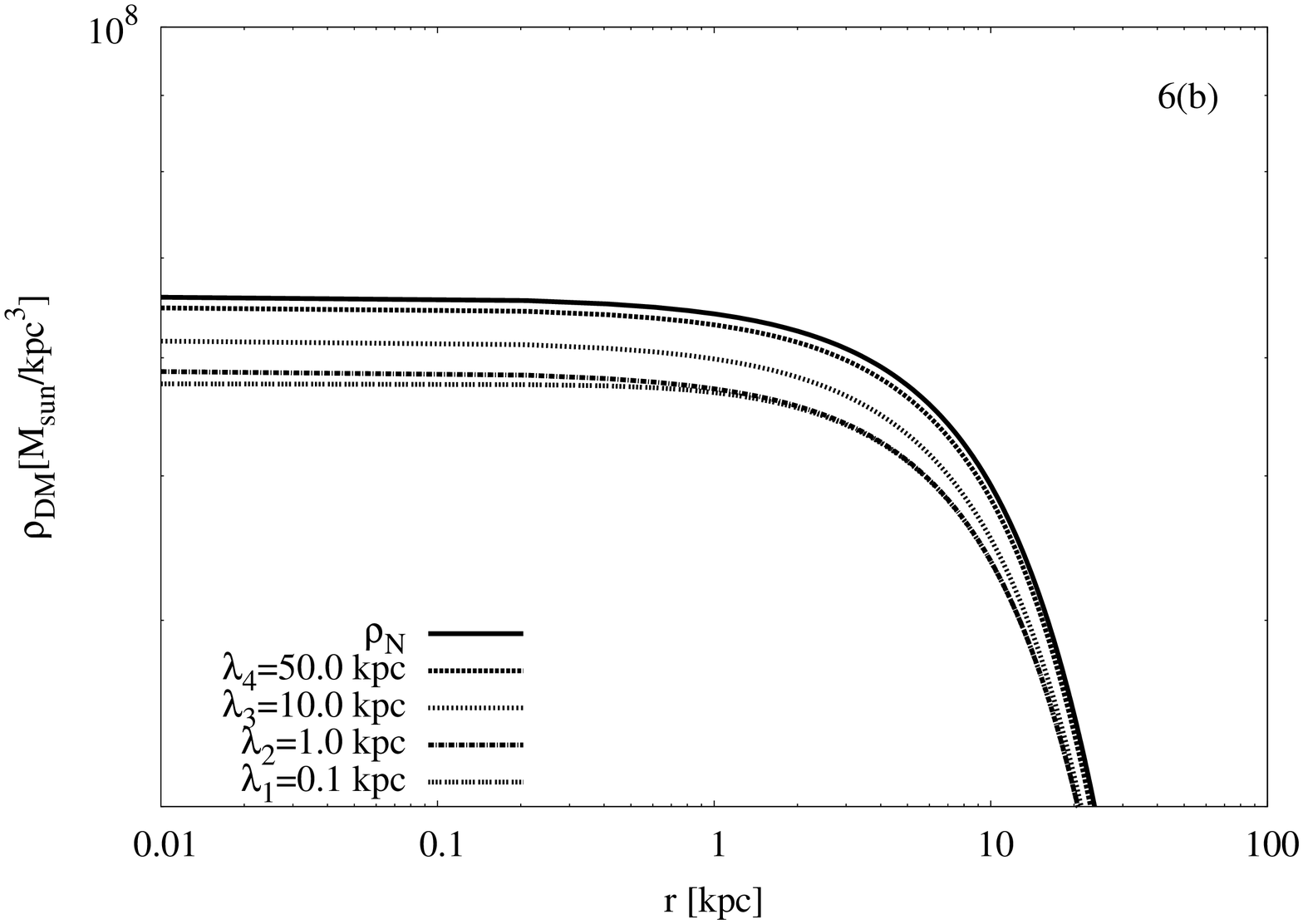} 
\caption{{\textbf (a)} It is shown the SF perturbation and {\textbf (b)} the DM profile for various $\lambda$ and fixed $\alpha=-0.1$.}
\label{phi-alpha-lambdas-n}
\end{figure*}

Figs. \ref{phi-alpha-lambdas-n}a-b show the SF and DM profiles for various 
$\lambda$, respectively.   The behaviors show systematic tendencies:  the smaller  
$\lambda$, the smaller  $| \delta \phi|$,  and again and for smaller $\lambda$, the decay is stronger in inner 
regions (Fig. \ref{phi-alpha-lambdas-n}a).  The SF complies with   
$\delta \phi <  (1+\alpha) c^{2} G_{N}^{-1}$, guaranteeing the validity of the perturbation approach.    On the other hand, the bigger  
 $\lambda$, the closer the DM profile approaches to the standard Newtonian 
 model (Fig. \ref{phi-alpha-lambdas-n}b).   

\begin{figure*}
\includegraphics[width=150mm]{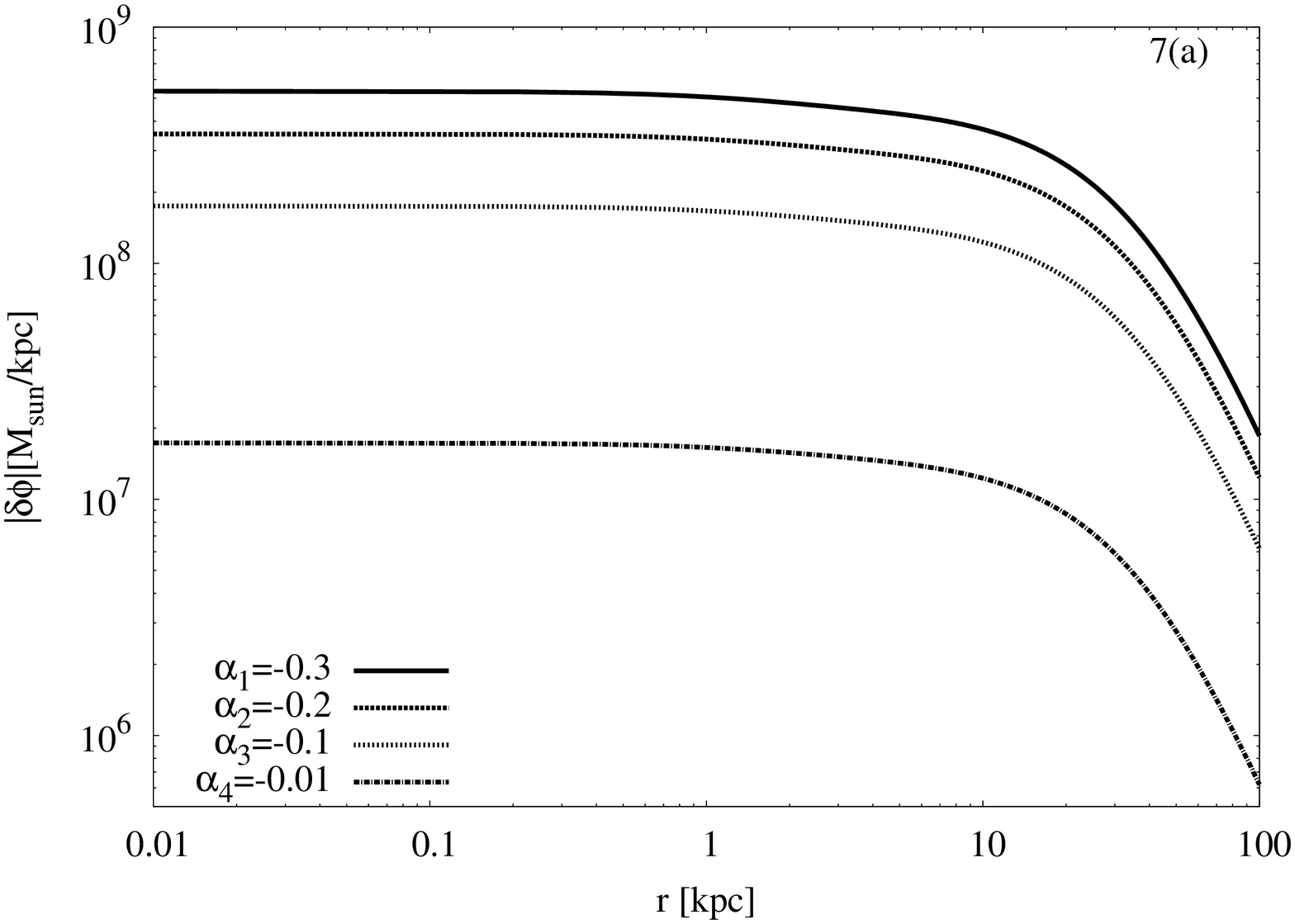} 
\includegraphics[width=150mm]{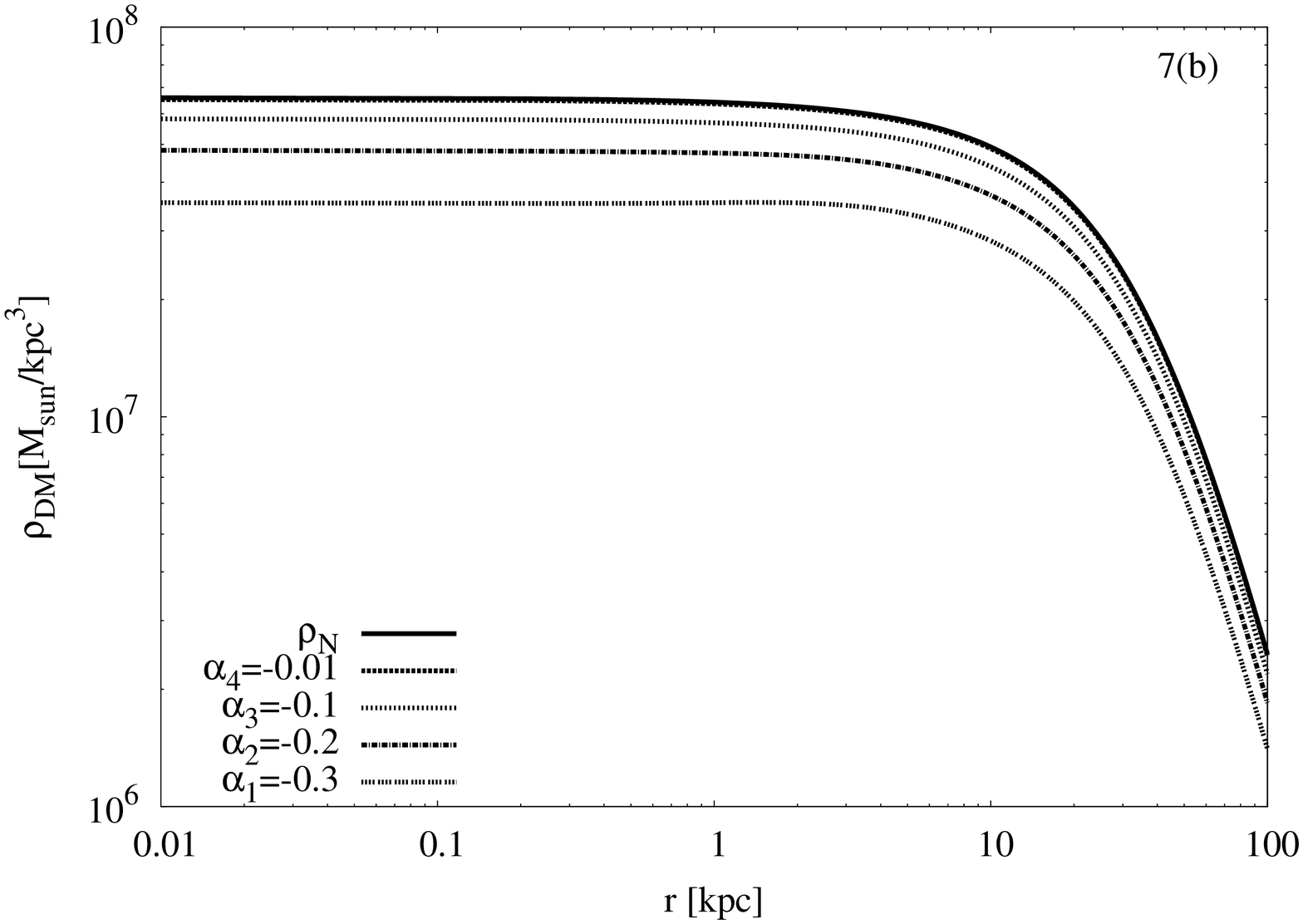}  
\caption{{\textbf (a)} It is shown the SF perturbation and {\textbf (b)} the DM profile for various negative $\alpha$ and fixed $\lambda=1$ kpc.}
\label{phi-lambda-alphas-n}
\end{figure*}

In Figs. \ref{phi-lambda-alphas-n}a-b we plotted the SF and DM profiles  for 
various negative $\alpha$ and fixed $\lambda = 1.0$ kpc.  Again the constraint on  
$\delta \phi $ is fulfilled.  As expected, for smaller  $|\alpha|$ the DM profile tends to the standard Newtonian one ($\alpha=0$).

\section{Discussion and conclusions}


We have considered a general STT in its weak energy limit in which two 
free parameters appear ($\lambda, \alpha$). This pair can be constrained    
by the above-mentioned observations: the first parameter is the Compton wavelength associated
with a light boson particle, which we have taken to be of the
order of kiloparsecs; this value implies a change of the Newtonian constant only at  
distances of the order of galactic scales or bigger, and therefore does not 
conflict with local deviations of Newtonian dynamics \cite{FiTa99}.  The other parameter is the
strength of the new scalar force, given by $\alpha$, which is
subject to cosmological constraints. Accordingly, we have taken values 
for $\alpha$ within the range $-0.3 \le \alpha \le 3$. We do not use bigger positive (negative) 
values for $\alpha$ because they predict a weaker (stronger) gravitational constant on
scales larger than $\lambda$, and bring  some un\-accep\-table  cosmological 
effects  \cite{UmIcYa05,NaChSu02,ShShYoSu05}.

Using  the STT formalism, we have constructed a ga\-lac\-tic model with a  distribution of stars and DM that  
obey a rotation profile compatible with observations given by the URC of \cite{Sa07}, as is shown 
in  Fig. \ref{rcs}.   This fitting determines the form of the effective gravitational potential, $\Phi_N$, given 
by Eq. (\ref{ph-N}).  Then, by taking  a typical density profile for baryons (Freeman exponential profile), the
only two quantities to be determined are the SF fluctuation and DM profile, which 
are given by Eqs. (\ref{phibar}) and (\ref{rho-dm}), respectively, for which we have found 
numerical solutions\footnote{In other models, such as the gravitational suppression 
hypothesis \cite{PiMa03} one assumes  a NFW DM profile to determine the parameters ($\lambda,\, \alpha$) of that theory by fitting 
theoretical curves to the rotation curves data \cite{FrSa07}.  In our approach we determine the DM profile by setting a fitted 
universal rotation curve that is generic for different spiral types \cite{Sa07}.}. In a 
 previous work \cite{CeRoNu07} we have found analytic 
solutions for a flat rotation curve profile. Those solutions are  similar to NFW's \cite{Na96-97}.  The 
numerical solutions found here are shallow near the galaxy center, since the URCH, Eq. (\ref{ph-rot-prof}), stems from the Burkert 
profile  \cite{Bu95}, which is shallow for $r\ll r_0$.   In order to make a comparison with the standard Newtonian model,  we turned off the 
SF and solved for the density in 
the standard Poisson using Eq. (\ref{ph-N}).  The resulting profile is called $ \rho_{N} $ and is the well known Burkert profile. 
The DM profile found here is slightly shallower (cuspier) than the Newtonian profile for positive (negative) values of $\alpha$, and 
in any case much shallower than NFW's.  
To quantitatively analyze the solutions we have considered the two scales involved in the URC fitting, that are the disc radius, $r_d$, and the 
DM scale, $r_0$.  The  DM solutions  near the galactic center ($r \ll r_d$) with $\alpha$ positive are a bit shallower 
($\rho \sim r^{-0.00006^{+0.00002}_{-0.00001}}$)  than those with $\alpha$ negative 
($\rho \sim r^{-0.00011^{+0.00004}_{-0.00002}}$), and 
both of them are much shallower than NFW's 
profiles \cite{Na96-97, Po03, Na04, Ha04}.  Solutions at $r \sim r_{d}$  
for  $\alpha$ positive decay as ($\rho_{DM} \sim r^{- 0.040^{+0.011}_{-0.007}}$) and for   $\alpha$ negative decay  
weaker  ($\rho \sim r^{-0.035^{+0.015}_{-0.009}}$).  At  $r\sim r_{0}$ for positive $\alpha$ the DM profile  
behaves as   $\rho \sim r^{-1.44^{+0.34}_{-0.26}}$ and for negative $\alpha$ the behavior is  
$\rho \sim r^{-1.44^{+0.34}_{-0.26}}$.  The
uncertainties stemming from the allowed values of the fitting parameters reported in table I.  For the allowed values of 
$r_0$, the DM exponents for $r \ll r_d$ vary between $20\%-34\% $, for $r \sim r_d$ between $17\%-29\% $, and  for 
$r \gg r_d$ vary  between $18\%-24\%$.   The slopes of the DM profiles  
change smoothly, so that their behavior does not quali\-ta\-tively differ from their mean value.  
With respect to the variations of $M_d$ and $\rho_0$ the exponents vary less than (or of the order of)  $1 \%$.       

On the other hand, for positive (negative)  
$\alpha$, $\rho_{DM} $ is bigger (smaller) than  $ \rho_{N} $ always, since the effect of  the SF is to 
diminish (augment) the effective gravitational 
constant for $r>\lambda$, being $G_{\rm eff}= G (1+\alpha e^{-r/\lambda})/(1+ \alpha) $, thus  
it is necessary to compensate it with a corresponding larger (smaller) DM density.   Finally, asymptotically, for 
$r \gg r_{0}$, the exterior solution is valid, and thus   $\rho_{\rm DM} \sim r^{-3}$ is similar to  
$\rho_{\rm NFW} \sim r^{-3}$. 

We have found numerical solutions for  the allowed parameter space for  strength of the SF potential,   
$-0.3 \le \alpha \le 3.0$, and within galactic distances,  $0.1 \, {\rm kpc} \le \lambda \le 50$ kpc.   The results indicate 
some systematic tendencies:  the effect of the SF is more apparent for 
$\lambda < r_{0}$, and its influence attenuates for $\lambda > r_{0}$, since in this regime the 
behavior is essentially Newtonian ($\lambda = \infty$) for $r< r_{0}$.   For small $|\alpha|$ the DM profile tends 
to the standard Newtonian one ($\alpha=0$).  

The intention of the present work was to study the influence of a massive SF in a galaxy model that is compatible with a typical baryon distribution 
and follows the URC of observed galaxies. This construction fixes the Newtonian potential, and for the STT parameters  
analyzed, the resulting DM profile is shallow at the center of the halo. These results are encouraging showing the important role that both 
contributions, DM and STT, could have in the dynamics of the systems.  We think that numerical N-body simulations using the 
STT of gravity have to be done to confirm the solutions discussed here.  Some  preliminary 
computations have been carried out  in  Ref. \cite{RoGoGaCe07,Ro08}. 

\begin{acknowledgments}
This work was supported by CONACYT Grant Nos. 84133-F, U47209-F, and I0101/131/07 C-234/07.  We thank P. Salucci and R. Kuzio de Naray and 
collaborators \cite{KuMcBl08} for providing us with the rotation curves data used here.  
\end{acknowledgments}

\end{document}